\documentstyle[aps,preprint]{revtex}
\begin{document}
\draft
\title{Fresh inflation with nonminimally coupled inflaton field}
\author{Mauricio Bellini\footnote{E-mail address: mbellini@mdp.edu.ar;
bellini@ginette.ifm.umich.mx}}
\address{Instituto de F\'{\i}sica y Matem\'aticas, \\
Universidad Michoacana de San Nicol\'as de Hidalgo, \\
AP:2-82, (58041) Morelia, Michoac\'an, M\'exico.}

\maketitle
\begin{abstract}
I study a fresh inflationary model with a scalar field nonminimally coupled
to gravity. An example is examined. I find that, as larger is the value
of $p$ ($a \sim t^p$), smaller (but larger in its absolute value) is the
necessary value of the coupling $\xi$ to the inflaton field fluctuations
can satisfy a scale invariant power spectrum.

\end{abstract}
\vskip 2cm
\noindent
{\rm Pacs:} 98.80.Cq \\
\vskip 2cm
\section{Introduction}
The inflationary model is one of
the most promising, for the early stage of the
universe in modern
cosmology\cite{1,2}. It not only gives a natural explanation for the horizon,
flatness, and monopole problems but also provides density perturbations
as seeds for the large scale structure in the universe.
The standard inflationary period
proceeds while a scalar field called an inflaton slowly evolves
along a sufficently flat potential.

The standard slow - roll inflation model separates expansion and reheating
into two distinguished time periods. It is first assumed that
exponential expansion from inflation places the universe in a supercooled
second order phase transition. Subsequently thereafter the universe is
reheated. Two outcomes arise from such a scenario. First,
the required density perturbations in this cold universe are left to be
created by the quantum fluctuations of the inflaton. Second, the
scalar field oscillates near the
minimum of its
effective potential and produces elementary particles. These
particles interact with each other and eventually they come
to a state of thermal equilibrium at some temperature $\theta$. This process
completes when all the energy of the classical scalar field
transfers to the thermal energy of elementary particles.
The temperature of the universe at this stage is called
the reheating temperature\cite{21}.
From a viewpoint of quantum field theory in curved
spacetime, it is natural to consider that the inflaton field
$\phi$ couples nonminimally to the spacetime curvature $R$ with
a coupling of $\xi R \phi^2/2$\cite{3}. For example, in the new inflationary
model\cite{lin}, the existence of nonminimal coupling prevents inflation
in some cases, because the flatness of the potential of inflation is
destroyed around $\phi = 0$. In the chaotic inflation model\cite{llin}, the
coupling $\xi$ in potentials like $V(\phi) = {m^2 \over 2} \phi^2$ and
$V(\phi) = {\lambda^2 \over 4} \phi^4$ is restricted to
$|\xi| < 10^{-3}$\cite{mae}.

Very recently a new model of inflation called {\em fresh inflation}
was proposed\cite{22,222,2222}. 
As in chaotic inflation\cite{llin}, in this model
the universe begins from an unstable primordial matter field
perturbation with energy density nearly $M^4_p$ and chaotic
initial conditions. Initially the universe there is no thermalized
[$\rho_r(t=t_0)=0$]. Later, the universe describes a second order
phase transition, and the inflaton rolls down slowly towards its minimum
energetic configuration.
Particles production and heating occur together during the inflationary
expansion of the universe, so that the radiation energy density
grows during fresh inflation ($\dot\rho_r >0$). The Yukawa interaction
between the inflaton field and other fields in a thermal bath
lead to dissipation which is responsible for the slow rolling of the
inflaton field. So, the slow-roll conditions are physically justified
and there are not a requirement of a nearly flat potential in fresh
inflation.
Furthermore,
there is no oscillation of the inflaton field around the minimum
of the effective
potential due to the strong dissipation produced by the Yukawa
interaction ($\Gamma \gg H$). This fact also provides thermal equilibrium
in the last stages of fresh inflation.

The aim of this paper is the study of nonminimal coupling of the inflaton
field in the fresh inflationary scenario. This topic will be studied in Sect.
II. In Sect. III, I study the dynamics of inflaton fluctuations for
$\xi \neq 0$ and, in Sect. IV, an example for this formalism is examined.
Finally, in Sect. V, some final comments are developed.

\section{Fresh Inflation with nonminimal coupling}

We study a model of fresh inflation
where an inflaton field $\phi$ is nonminimally coupled
with a scalar curvature $R$
\begin{equation}\label{1}
{\cal L} = \sqrt{-g} \left[\frac{R}{16\pi G} - \frac{1}{2}\left(
\nabla \phi\right)^2 - V(\phi) -\frac{1}{2} \xi R \phi^2+{\cal L}_{int}\right],
\end{equation}
where $G=M^{-2}_p$ is the gravitational constant, $M_p=1.2 \  {\rm GeV}$ 
is the Planckian mass and ${\cal L}_{int} = -{\rm g}^2 \phi^2 \psi^2$ takes
into account the interaction between the inflaton and the scalar
field $\psi$.
The Lagrangian (\ref{1}) can be rewritten as
\begin{equation}\label{2}
{\cal L} = \sqrt{-g} \left[\frac{R}{16\pi G_{eff}(\phi)} - \frac{1}{2}\left(
\nabla \phi\right)^2 - V(\phi)+{\cal L}_{int}\right],
\end{equation}
where $G_{eff}(\phi)={G \over 1-\phi^2/\phi^2_c}$ with 
$\phi^2_c = {M^2_p \over 8\pi \xi}$ and $R= 6(a \ddot a+\dot a^2)/a^2$,
($a$ is the scale factor of the universe).
In order to connect to our
present universe, $G_{eff}$ needs to be positive so that in the case
$\xi >0$ we require $\phi^2 < {M^2_p \over 8\pi |\xi|}$.
In this paper I study only the case $\xi \neq 0$. The case $\xi=0$ for 
a minimally coupled scalar field in fresh inflation 
was analyzed in\cite{22}.

The Einstein equations for a globally flat, isotropic and homogeneous
universe described by a Friedmann-Robertson-Walker metric
$ds^2 = -dt^2 + a^2(t) dr^2$, are given by
\begin{eqnarray}
3 H^2 & =& 8\pi G\left[\left(\frac{1}{2}-2\xi\right) \dot\phi^2 + V(\phi) +
\rho_r\right], \label{4} \\
3H^2 + 2 \dot H & = & -8\pi G\left[\left(\frac{1}{2}-2\xi\right)
\dot\phi^2-V(\phi) + \rho_r\right], \label{5}
\end{eqnarray}
where $H={\dot a\over a}$ is the Hubble parameter and $a$ is the
scale factor of the universe. 
The overdot denotes
the derivative with respect to the time.
On the other hand, if $\delta=\dot\rho_r+4H\rho_r$
describes the interaction between the inflaton and the bath,
the equations of motion for $\phi$ and $\rho_r$
are
\begin{eqnarray}
&& \ddot\phi+3H\dot\phi+V'(\phi) + \xi R \phi
+ \frac{\delta}{\dot\phi}=0,\label{6}\\
&& \dot\rho_r+4H\rho_r-\delta=0. \label{7}
\end{eqnarray}
As in previous papers\cite{22,222}, I will consider a Yukawa interaction
$\delta = \Gamma(\theta) \  \dot\phi^2$, where $\Gamma(\theta)=
{g^4_{eff}\over 192\pi}\theta$\cite{41} and $\theta \sim \rho^{1/4}_r$ 
is the temperature of the bath.
Slow-roll conditions must be imposed to assure nearly de Sitter solutions
for an amount of time, which must be long enough to solve the 
flatness and horizon problems. If $p_t={\dot\phi^2 \over 2}+{\rho_r\over 3} - V(\phi)$
is the total pressure and $\rho_t=\rho_r+{\dot\phi^2 \over 2}+V(\phi)$ is the
total energy density, hence the parameter $F={p_t+\rho_t \over \rho_t}$
which describes the evolution of the universe during inflation\cite{5}, is
\begin{equation}\label{8}
F= - \frac{2\dot H}{3 H^2} = \frac{\left(1-4\xi\right)
\dot\phi^2+\frac{4}{3} \rho_r}{\rho_r+ \left(\frac{1}{2}-2\xi\right)
\dot\phi^2+V} >0.
\end{equation}
When fresh inflation starts (at $t=G^{1/2}$), the radiation energy 
density is zero, so that $F\ll 1$. 

From the
two equalities in eq. (\ref{8}) one obtains the following equations
\begin{eqnarray}
&& \dot\phi^2  \left[(2-F)\left(\frac{1}{2}+2\xi\right)\right]
+ \rho_r\left(\frac{4}{3}-F\right)-F \  V(\phi)=0, \label{9} \\
&& H= \frac{2}{3 \int F \  dt}.\label{10}
\end{eqnarray}
Furthermore, due to $\dot H = H'\dot\phi$ (here the prime denotes
the derivative with respect to the field), from que first equality
in eq. (\ref{8}) it is possible to obtain the equation that
describes the evolution for $\phi$
\begin{equation}\label{11}
\dot\phi=-\frac{3 H^2}{2 H'}F,
\end{equation}
and replacing eq. (\ref{11}) in (\ref{9}) the radiation energy
density can be described as functions of $V$, $H$ and $F$\cite{22}
\begin{equation}\label{12}
\rho_r = \left(\frac{3F}{4-3F}\right) V - \frac{27}{4}
\left(\frac{H^2}{H'}\right)^2 
F^2\left[\frac{(2-F)}{(4-3F)} \left(\frac{1}{2}+2\xi\right)\right]
\end{equation}
Finally, replacing (\ref{11}) and (\ref{12}) in eq. (\ref{4}),
the potential can be written as
\begin{equation}\label{13}
V(\phi) = \frac{3(1-8\pi\xi G\phi^2)}{8\pi G} 
\left[\left(\frac{4-3F}{4}\right)
H^2 + \frac{3\pi}{2} \frac{G F^2}{(1-8\pi\xi G \phi^2)}
\left(\frac{H^2}{H'}\right)^2 (1+8\xi)\right].
\end{equation}

Fresh inflation was proposed for a global group $O(n)$, involving
a single $n$-vector multiplet of scalar fields $\phi_i$\cite{6},
such that making $(\phi_i\phi_i)^{1/2}\equiv \phi$, the effective
potential $V_{eff}(\phi,\theta)=V(\phi)+\rho_r(\phi,\theta)$ can be
written as
\begin{equation}\label{14}
V_{eff}(\phi,\theta) = \frac{{\cal M}^2(\theta)}{2} \phi^2+
\frac{\lambda^2}{4}\phi^4,
\end{equation}
where ${\cal M}^2(\theta) = {\cal M}^2(0)+{(n+2) \over 12} \lambda^2\theta^2$
and $V(\phi)= {{\cal M}^2(0)\over 2} \phi^2+{\lambda^2 \over 4}\phi^4$.
Furthermore ${\cal M}^2(0) >0$ is the squared mass at
zero temperature, which is given by ${\cal M}^2_0$ plus renormalization
counterterms in the potential ${1 \over 2} {\cal M}^2_0 (\phi_i\phi_i)+
{1 \over 4} \lambda^2 (\phi_i\phi_i)^2$\cite{6}.
I will take 
into account the case without symmetry breaking ${\cal M}^2(\theta) >0$
for any temperature $\theta$, so that
the excitation spectrum consists of $n$ bosons with
mass ${\cal M}(\theta)$.
Note that the effective potential (\ref{14}) 
is invariant under $\phi \rightarrow -\phi$ 
reflexions and $n$ is the number of created particles due to
the interaction of $\phi$ with the particles in the thermal bath, such
that\cite{22}
\begin{equation}\label{15}
(n+2) = \frac{2\pi^2}{5\lambda^2}g_{eff} \frac{\theta^2}{\phi^2},
\end{equation}
because the radiation energy density is given by $\rho_r={\pi^2 \over 30}
g_{eff} \theta^4$  ($g_{eff}$ denotes the effective degrees
of freedom of the particles and it is assumed that $\psi$ has no                                                
self-interaction).\\

\section{Dynamics of the inflaton field and power spectrum
of the fluctuations}

In this section I will study the dynamics of the inflaton field to make
an estimation for the energy density fluctuations in a spatially flat
Friedmann-Robertson-Walker (FRW) metric
\begin{equation}
ds^2= -dt^2 + a^2(t) dx^2.
\end{equation}
The dynamics for the spatially homogeneous inflaton field $\phi$ is given
by
\begin{equation}\label{cl}
\ddot\phi + (3 H +\Gamma)\dot\phi + \xi R \phi + V'(\phi) =0,
\end{equation}
where $V'(\phi)\equiv {dV\over d\phi}$. The term $\Gamma \dot\phi$ is added
in the scalar field equation of motion (\ref{cl}) to describe the
continuous energy transfered from $\phi$ to the thermal bath. This persistent
thermal contact during fresh inflation is so finely adjusted that the scalar
evolves always in a damped regime.

Furthermore, the fluctuations $\delta\phi(\vec x,t)$ are described by the
equation of motion
\begin{equation}\label{q}
\ddot\delta\phi - \frac{1}{a^2} \nabla^2\delta\phi +
(3H + \Gamma) \dot\delta\phi + \left[\xi R+ V''(\phi)\right]\delta\phi=0,
\end{equation}
where $R=12 H^2 + 6 \dot H$.
Here, the additional second term appears because the fluctuations $\delta\phi$
are spatially inhomogeneous.
The equation
for the modes
$\chi_k(\vec x,t) = \xi_k(t) e^{i\vec k.\vec x}$ and
$\chi^{*}_k(\vec x,t)=
e^{-i\vec k.\vec x} \xi^*_k(t)$,
of redefined fluctuations
$\chi=a^{3/2} e^{{1\over 2}\int \Gamma dt} \  \delta \phi $
(which can be written as a Fourier expansion as)
\begin{equation}
\chi(\vec x,t) = \frac{1}{(2\pi)^{3/2}} {\Large\int} d^3k
\left[a_k \  \chi_k(\vec x,t)
+ a^{\dagger}_k \  \chi^*_k(\vec x,t)\right],
\end{equation}
is
\begin{equation}\label{xi}
\ddot\xi_k + \omega^2_k \xi_k=0,
\end{equation}
where $\omega^2_k = a^{-2}\left[k^2 - k^2_0\right]$ is the
squared frequency for each mode and
$k^2_0$ is given by 
\begin{equation}\label{k^2}
k^2_0(t) = a^2 \left\{\frac{9}{4}\left(H+\Gamma/3\right)^2 -12\xi H^2 +
3\left[\left(1-2\xi\right)\dot H+\dot\Gamma/3\right]
- \left[12\xi H^2+V''[\phi(t)]\right]\right\}.
\end{equation}
Here,
the time-dependent wave number $k_0(t)$ separates the infrared
(IR) and ultraviolet (UV) sectors. The IR sector includes the
long wavelength modes ($k < k_0$) and the UV sector takes into
account the short wavelength modes ($k > k_0$).
Furthermore
($a_k$, $a^{\dagger}_k$) are respectively the annihilation and
creation operators, which complies with the commutation relations
$[a_k,a^{\dagger}_{k'}] = \delta^{(3)}(\vec k - \vec{k'})$ and
$[a^{\dagger}_k,a^{\dagger}_{k'}]=[a_k, a_{k'}]=0$.
If we take
$\xi_k = \xi^{(0)}_k e^{\int g dt}$, the equation for $\xi^{(0)}_k$
can be approximated to
\begin{equation}\label{eq}
\ddot\xi^{(0)}_k + a^{-2}\left[k^2-\tilde{ k}^2_0 \right]\xi^{(0)}_k=0,
\end{equation}
where $\tilde{k}^2_0$ is given by
\begin{equation}\label{til}
\tilde{k}^2_0 = a^2 \left\{ H^2 \left( \frac{9}{4}-12\xi \right)
+ 3 \left(1-2\xi\right) \dot H
+ \dot\Gamma- V''[\phi(t)]\right\}.
\end{equation}
The function
$g(t)$ only takes into account the thermal effects. The
differential equation for $g$ is
\begin{equation}
g^2 + \dot g = \frac{3}{2} H\Gamma + \frac{1}{4} \Gamma^2,
\end{equation}
with initial condition $g(t=t_0)=0$, since the temperature when fresh
inflation starts is zero.
The squared fluctuations for super Hubble scales ($k^2 \ll k^2_0$),
are given by
\begin{equation}\label{sf}
\left<\left(\delta\phi\right)^2\right> = \frac{a^{-3}}{2\pi^2}
{\cal F}(t) {\Large\int}^{k_0(t)}_{0}
dk \  k^2 \xi^{(0)}_k \left(\xi^{(0)}_k\right)^*,
\end{equation}
where the asterisk denotes the complex conjugate and the function ${\cal F}$ is
given by ${\cal F}(t)=e^{\int\left[ 2g(t)-\Gamma\right] \  dt}$. 

\section{An example}

We consider the case in which $F$ is a constant during
inflation and the Hubble parameter is given by
\begin{equation}\label{H}
H(\phi) = 4 {\cal M}(0) \sqrt{\frac{\pi G}{3(4-3 F)}} \  \phi,
\end{equation}
From the eq. (\ref{11}) one
obtains the time dependence for the inflaton field
\begin{equation}
\phi(t) = \frac{1}{6 F {\cal M}(0)} \sqrt{\frac{3(4-3 F)}{\pi G}} \  t^{-1}.
\end{equation}
This is a very interesting case because $\phi(t)$ never
holds zero, as in the model initially proposed for fresh
inflation\cite{22}.

The initial value of $\phi$ being given by the equation
$\rho_r(\phi_i)=0$:
\begin{equation}
\phi_i = \sqrt{\frac{4-3F}{\pi G\left[F(36\xi+12)+16\xi-F^2(9+33\xi)\right]}}.
\end{equation}
Furthermore, since $V(\phi) =
\left({\cal M}^2(0)/2\right) \phi^2 + \left(\lambda^2/4\right) \phi^4$,
replacing eq. (\ref{H}) in eq. (\ref{13}), one obtains
the expression for $\lambda$ as a function of ${\cal M}(0)$
\begin{equation}
\lambda^2 = \frac{16 \pi G}{3(4-3 F)}\left[
\frac{9}{4} F^2 (1+3\xi)-3\xi(4-3 F)\right] {\cal M}^2(0),
\end{equation}
so that the scalar potential can be written as
\begin{equation}
V(\phi) = \frac{{\cal M}^2(0)}{2} \phi^2 + \frac{4\pi G}{3(4-3F)}
\left[ \frac{9}{4} F^2 (1+3\xi) - 3\xi (4-3F)\right] {\cal M}^2(0) \phi^4.
\end{equation}
In this paper I will consider ${\cal M}^2(0) \simeq 10^{-12} \  {\rm G^{-1}}$.

Notice that for $\xi =0$ one obtains $\lambda^2 = {12\pi G F^2 \over
(4-3 F)} {\cal M}^2(0)$\cite{22}.
The scale factor evolves as $a \sim t^{2/(3F)}$ and the temperature
is given by the equation [see eq. (\ref{7}) where
$\delta =\Gamma(\theta) \dot\phi^2$, $\Gamma(\theta) =
{g^4_{eff}\over 192 \pi} \theta(t)$ and $\rho_r$ is given by
eq. (\ref{12})]
\begin{equation}
\theta(t) = \frac{192 \pi}{g^4_{eff}} \frac{1}{\dot\phi^2}
\left[\dot\rho_r + 4 H \rho_r\right],
\end{equation}
which, for $t \gg G^{1/2}$ gives
\begin{equation}
\theta(t) \simeq \frac{768 \pi {\cal M}^2(0)}{(4-3 F) g^4_{eff}} \  t.
\end{equation}
The number of created particles for $g_{eff} \simeq 10^2$ is
[see eq. (\ref{15})] 
\begin{equation}
n\simeq 2.2 \times 10^{-7} \frac{{\cal M}^4(0)}{(4-3F)^2} t^4,
\end{equation}
which increases with time. So, during fresh inflation the expansion is
accompanied by intense particle creation.
Hence, the decay width of the inflaton field for $F\ll 1$ is
\begin{equation}
\Gamma[\theta(t)] \simeq {\cal M}^2(0) \  t.
\end{equation}
To obtain ${\tilde{k}^2_0 \over a^2}$ in eq. (\ref{til}), we make $p = 2/(3F)$
\begin{equation}
\frac{\tilde{k}^2_0}{a^2} =
\left[\left(\frac{9}{4}+24\xi\right)p^2-3\left(1+4\xi\right)p-
3\left(1+3\xi\right)\right] t^{-2}.
\end{equation}
Hence, the equation for $\xi^{(0)}_k(t)$ is
\begin{equation}
\ddot\xi^{(0)}_k + \left\{\frac{k^2 t^{-2p}}{a^2_0 t^{-2p}_0} -
\left[\left(\frac{9}{4} +24 \xi\right)p^2-3\left(1+4\xi\right)p-
3\left(1+3\xi\right)\right] t^{-2} \right\} \xi^{(0)}_k=0.
\end{equation}
The general solution (for $\nu \neq 0,1,2,...$) is
\begin{equation}
\xi^{(0)}_k(t) = C_1 \sqrt{\frac{t}{t_0}} {\cal H}^{(1)}_{\nu}
\left[\frac{k t^{1-p}}{a_0 t^{-p}_0 (p-1)}\right]
+ C_2 \sqrt{\frac{t}{t_0}} {\cal H}^{(2)}_{\nu}
\left[\frac{k t^{1-p}}{a_0 t^{-p}_0 (p-1)}\right],
\end{equation}
where $\nu = \sqrt{(9+96\xi)p^2 - 12(1+4\xi)p -
(11+36\xi)}/[2(p-1)]$, which, for a given value of $\xi$, tends to
a constant as $p \rightarrow \infty$. For $\xi = 0$ $\nu \rightarrow 3/2$
as $p \rightarrow \infty$.
Furthermore, $({\cal H}^{(1)}_{\nu}, {\cal H}^{(2)}_{\nu})$ are the
Hankel functions. These functions take the small-argument
limit $\left.{\cal H}^{(2,1)}_{\nu}[x]\right|_{x\ll 1} \simeq
{(x/2)^{\nu} \over \Gamma(1+\nu)} \pm {i\over \pi}
\Gamma(\nu) \left(x/2\right)^{-\nu}$.
We can take the Bunch-Davis vacuum such that $C_1=0$ and
$C_2 = \sqrt{\pi/2}$\cite{BD}. 
Notice that $\xi^{(0)}_k$ is the solution for the
modes when the interaction is negligible ($\Gamma \propto \theta \simeq 0$).
The function $g(t)$
only takes into account the thermal effects.
Taking into account the small-argument limit for the Hankel functions
and the Bunch-Davis vacuum, we obtain
\begin{equation}
\xi^{(0)}_k \left(\xi^{(0)}_k\right)^*
\simeq \frac{2^{2\nu}}{\pi^2} \Gamma^2(\nu) \left[
\frac{a_o (p-1)}{t^p_0} t^{(p-1)} \right]^{2\nu} k^{-2\nu},
\end{equation}
so that the integral controlling the presence of infrared divergences
in eq. (\ref{sf}) is
${\Large\int}^{k_0(t)}_0 dk \  k^{2(1-\nu)}$, with
a power spectrum ${\cal P}_{<(\delta\phi)^2>} \sim k^{3-2\nu}$. Hence, 
the condition $n_s=3/2-\nu$ gives a spectral index $n_s \simeq 1$ according
with the experimental data\cite{8} for $\nu \simeq 1/2$. This implies
$p \simeq 2$ for $\xi =0$.
On the other hand, the condition $N =
\int^{10^{13} G^{1/2}}_{G^{1/2}} H(t) \  dt \ge 60$ (that implies $F\le 1/3$)
assures the solution of the horizon problem to give a sufficiently globally
flat universe. This condition implies $p \ge 2$.
On the other hand,
for $1/2 \le \nu <3/2$ there is no infrared divergence.
These conditions implies $\xi \le 0.3174$ and $p \ge 2$. The experimental
data\cite{8}
obtained from BOOMERANG-98, MAXIMA-1 and COBE DMR, is
consistent with $\nu \simeq 1/2$ to obtain a spectral index $n_s \simeq 1$.
Such a condition constrains the possible values for $p$ and $\xi$ to
$\xi \le 0$ and $p \ge 2$.

\section{Final Comments}

We have investigated the dynamics of a fresh inflationary scenario with
a inflaton field nonminimally coupled to gravity.
Fresh inflation attempts to build a bridge between the standard
and warm inflationary models, beginning from chaotic initial conditions
which provides naturality. In this sense, this model can be viewed as an
unification of both chaotic\cite{llin} and warm inflation\cite{wi} scenarios.
In our study the
inflaton field coupled nonminimally to a spacetime curvature $R$ by
means of an additional term ($-\xi R \phi^2/2$) in the Lagrangian.
In the example here studied, I find that the possible values for the
coupling are restricted to $\xi \le 0$, for $p \ge 2$. These values
becomes from experimental data obtained from BOOMERANG-98, MAXIMA-1 and
COBE DMR, which are consistent with a spectral index $n_s \simeq 1$, related
to $\nu \simeq 1/2$ in the example here studied. The most interesting here,
is that, as larger is the value of $p$ ($a \sim t^p$), smaller (but
larger in its absolute value because the permitted values
of $\xi$ are negative) is the value of $\xi$ necessary to satisfy
$\nu \simeq 1/2$, which is consistent with a scale invariant power spectrum
($n_s \simeq 1$) for the inflaton fluctuations.

\end{document}